\documentclass[aip,cha,reprint,amsmath,amssymb]{revtex4-2}

\usepackage{graphicx}
\usepackage{bm}
\usepackage[colorlinks=true,citecolor=blue,linkcolor=blue,urlcolor=blue]{hyperref}
\hypersetup{pdftitle={Same worst case, different risk: how transient response shapes failure probability},pdfauthor={Surachate Limkumnerd}}

\newcommand{\Pfail}{P_{\rm fail}}
\newcommand{\Plin}{P_{\rm lin}}

\begin{document}

\title{Same worst case, different risk: how transient response shapes failure probability}

\author{Surachate Limkumnerd}
\email{surachate.l@chula.ac.th}
\affiliation{Department of Physics, Faculty of Science, Chulalongkorn University, Bangkok 10330, Thailand}

\date{\today}

\begin{abstract}
The smallest disturbance that can push a dynamical system into failure is a standard measure of stability, but it does not determine how likely failure is under randomly directed disturbances of a given size.
The geometry of the transient response connects the two.
In the linearized dynamics, each failure boundary and each time after the disturbance define a response vector along the most damaging disturbance.
The convex hull of these vectors, the response hull, determines the linear failure-probability curve through its support function.
The longest vector sets the worst-case threshold. The mean and spread of the support function set where the curve rises and how steeply, but the mean does not rank systems by risk in general.
Under mild regularity conditions, an exact linear-response relation shows how a parameter change shifts the disturbance size at which a given fraction fails: through the mean susceptibility of the directions on that threshold.
For weakly damped oscillators, the susceptibility to damping is approximately proportional to the time of a direction's dominant response.
Damping thus acts mainly on late failure routes rather than on the early one that sets the worst case, changing the failure probability substantially while barely moving the worst-case threshold.
At fixed power imbalance, lowering the inertia instead nearly rescales the hull, moving worst case and risk together.
In benchmark power-grid models, the linear predictions agree closely with nonlinear simulations.
Strong nonlinear coupling, invisible to linear response, can break this picture; we show how it enters at second order.
\end{abstract}

\maketitle

\begin{quotation}
Two systems can have the same worst-case stability threshold yet very different probabilities of failure under random disturbances.
The reason lies in their transient responses: as a system evolves, different disturbance directions become dangerous at different times.
A worst-case calculation finds only the most dangerous of these failure routes and misses the others.
We show how the linear response, followed over time, determines the probability of crossing a failure boundary, and how a change in the response predicts the change in that probability.
Power grids illustrate the mechanism.
After a sudden disturbance, the generators swing back and forth.
The first swing can set the smallest disturbance that overloads a transmission line, while later swings make further disturbance directions dangerous.
Damping suppresses these later swings, so it lowers the probability of overload with little change in the worst case.
Lowering the inertia, by contrast, enlarges every swing alike, and the worst case registers it.
More generally, how the probability of failure responds to a change of the system is set by the disturbances that sit right at the failure threshold.
\end{quotation}

% =============================================================================
\section{Introduction}
\label{sec:intro}

How large a disturbance can a system survive?
For flows, structures, and power grids the standard answer is a worst case.
Hydrodynamic stability theory computes the optimal perturbation, the initial condition that grows most before it decays~\cite{ButlerFarrell1992,Trefethen1993Hydrodynamic,Schmid2007Nonmodal}.
Studies of transition to turbulence compute the minimal seed, the smallest disturbance that triggers turbulence~\cite{PringleKerswell2010MinimalSeed,Kerswell2018}.
Structural reliability computes the most probable failure point~\cite{HasoferLind1974FORM,DitlevsenMadsen1996Reliability}.
Power-grid studies compute the smallest disturbance that desynchronizes the grid~\cite{Eaves2025}.
All of these ask the same question: what is the smallest push that can cause failure?

Real disturbances, however, are not chosen by an adversary.
A power grid is hit by load jumps, faults, and fluctuations of many shapes, and a flow is stirred by noise of many forms.
What matters in practice is therefore a probability.
We model a disturbance as a kick, an instantaneous change of the system's state.
If a kick of a given size arrives in a random direction, how likely is it to cause failure?
We call this probability, as a function of disturbance size, the failure-probability curve, and refer to it loosely as the risk.
Such probabilities can be estimated by brute force, running many nonlinear simulations from random initial conditions.
This is the basis of basin stability and survivability in power grids~\cite{Menck2013BasinStability,Hellmann2016Survivability} and of probabilistic studies of transition in shear flows~\cite{Pershin2020}.
These estimates are reliable but expensive, and they do not explain why one system is riskier than another.

\begin{figure*}[t]
  \centering
  \includegraphics[width=\textwidth]{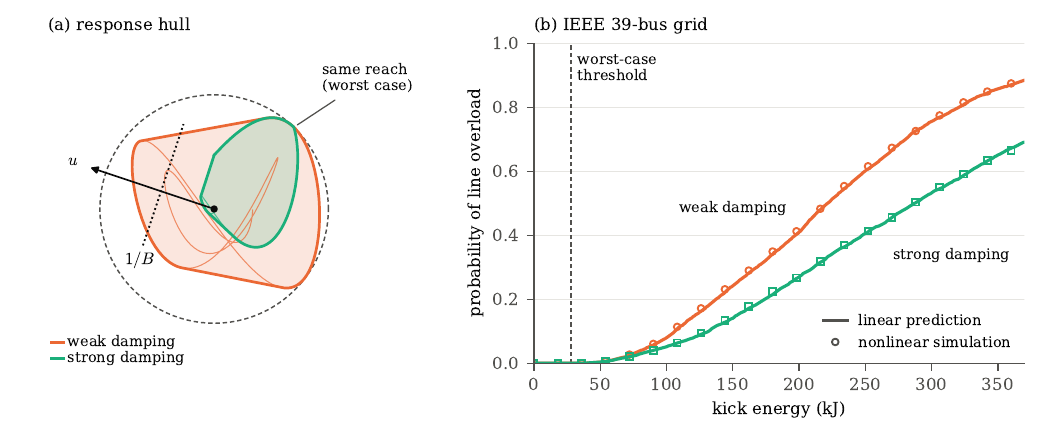}
  \caption{Same worst case, different risk.
  (a)~Schematic.
  The response vector of a damped system with two oscillation modes traces a curve (thin line).
  Its convex hull, together with the origin, is the response hull (shaded).
  To linear order, a kick of size $B$ in direction $u$ (arrow) causes failure if the hull crosses the line perpendicular to $u$ at distance $1/B$ from the origin (dotted).
  Weak and strong damping give the same reach, which is the worst case (dashed circle).
  Weak damping, however, leaves later swings that point in other directions, so its hull is wider and more kick directions are dangerous.
  (b)~Probability that a random frequency kick overloads a transmission line in the IEEE 39-bus test grid, for weak ($D=2$) and strong ($D=8$) generator damping.
  Line ratings are rescaled by less than 1.2\% so that both grids have the same worst-case threshold, 28~kJ (dashed line).
  Lines: linear prediction from the response hull; symbols: nonlinear simulations over 4096 random kick directions.}
  \label{fig:concept}
\end{figure*}

Is the worst case enough to predict the risk?
Figure~\ref{fig:concept}(b) shows that it is not.
It compares two versions of the IEEE 39-bus test grid~\cite{Athay1979,Pai1989} that differ only in how strongly their generators are damped.
The line ratings are adjusted by about 1\% so that both versions have the same worst case: in both, the smallest kick that can overload a line carries 28~kJ of kinetic energy.
Yet for kicks of 100 to 250~kJ, the weakly damped grid is 1.5 to 1.9 times more likely to overload a line.
(Kicks of this size change the generator speeds by 0.1 to 0.16~Hz in root mean square.)
The worst case marks where the risk begins, but it says little about where the risk becomes large.

In this paper we explain where the difference comes from and how to predict it from the linear response alone.
The explanation is geometric.
Follow every possible kick through the linearized dynamics.
At each moment and for each failure boundary, there is one kick direction that pushes the system toward that boundary most effectively.
We represent it by a response vector that points along this direction and whose length measures how effective the push is.
Together these response vectors form a cloud, and its convex hull, which we call the response hull, decides the linear risk [Fig.~\ref{fig:concept}(a)].
A kick is dangerous if the hull reaches far enough in the kick's direction.
The hull's largest radius, its reach, is the worst case.
But the chance that a random kick is dangerous depends on how far the hull reaches in a typical direction, its mean reach.
A response that turns over time, for example through successive swings that load a line from different directions, widens the hull without extending its reach.
Damping removes the later swings and so narrows the hull.

The mathematics behind this picture is classical.
Failure is decided by the largest of many linear functions of the kick, one for each line and each moment, and the probability that such a maximum exceeds a level has long been studied, both in the theory of random processes~\cite{Rice1944,Hotelling1939,AdlerTaylor2007} and in time-variant reliability analysis~\cite{Ditlevsen1983,HagenTvedt1991,AndrieuRenaud2004,Sudret2008}.
Statistical studies of transient growth~\cite{FrameTowne2024BeyondOptimal} come closer to the dynamics, but they ask how much a random disturbance grows, not how likely it is to cross a failure boundary.
What has been missing is the link between the two: which features of a transient response decide the risk, and how a change of the dynamics changes it.

Supplying that link is the aim of this paper, and power grids are our main example.
We find that when the damping or inertia of a grid changes but its worst case does not, the whole failure-probability curve shifts by the change in the hull's mean reach.
Damping, in particular, discounts each failure route, the line and moment at which a kick causes failure, according to when it occurs.
The risk therefore splits into two parts: failures during the first swing of the generators, which the worst case governs, and failures at later swings, which carry most of the effect of damping and become possible only above a second threshold.
The same discount gives an approximate law for the whole curve: when the damping changes, the kick energy at which a given fraction of kicks fails moves, relative to the worst case, in proportion to how much later than the worst case those kicks fail.
Inertia behaves differently: at a fixed power imbalance it rescales the whole hull, so the worst case tracks the risk.
These results hold also when a line fails only if it stays overloaded for a while.

The linear picture has a limit.
When the nonlinearity is strong, systems with the same linear response can differ in risk, because the linear response cannot see how the nonlinear terms couple the variables.
A second-order calculation shows how this coupling enters.

The paper is organized as follows.
Section~\ref{sec:linear} defines the failure-probability curve and its linear prediction, and tests the prediction on two benchmark grids.
Section~\ref{sec:hull} develops the response hull, three simple rules that follow from it, and what the geometry can and cannot guarantee.
Section~\ref{sec:grids} looks inside the grids: it locates the failures that damping changes, derives how damping moves the whole curve, and contrasts damping with inertia.
Section~\ref{sec:nonlinear} shows where the linear picture fails, and what a second-order calculation can and cannot correct.
Section~\ref{sec:discussion} discusses related work, practical use, and limitations.

% =============================================================================
\section{Predicting failure from the linear response}
\label{sec:linear}

\subsection{The failure-probability curve}

Consider a system resting in a stable state.
It fails if some quantity crosses a limit, for example if the power flowing through any transmission line exceeds the line's rating.
At time zero the system receives a kick of size $B$ in a direction $u$, and we ask whether any limit is crossed within a fixed time window.

For a power grid, a natural kick is a sudden change in the rotation speeds of the generators, such as a short power imbalance would cause.
Because the speeds set the electrical frequency of each machine, we call it a frequency kick.
We measure its size $B$ so that the kinetic energy it carries is proportional to $B^2$; for the grids below, $E=B^2\times265$~kJ.
The direction $u$ is drawn uniformly with respect to the kick energy, so that all kicks of the same energy are equally likely (Appendix~\ref{app:grids}).

The failure-probability curve $\Pfail(B)$ is the fraction of directions that lead to failure.
It rises from zero for small kicks toward one for large kicks, and its shape is what this paper is about.

\subsection{Linear prediction}

For small kicks the system responds linearly.
The change in the flow on line $b$ at time $t$ is then proportional to the kick, $B\,\bm{\ell}_b(t)\cdot u$, where the vector $\bm{\ell}_b(t)$ is obtained from the linearized dynamics.
Line $b$ fails when this change exceeds its margin $m_b$, the distance between the present flow and the rating.
It is convenient to divide by the margin and define the response vector
\begin{equation}
  \bm{r}_b(t) = \frac{\bm{\ell}_b(t)}{m_b}.
  \label{eq:r}
\end{equation}
A kick in direction $u$ then causes failure if $B\,\bm{r}_b(t)\cdot u \ge 1$ for some line $b$ and some time $t$.
All lines and times together give one number per direction,
\begin{equation}
  g(u) = \max\Big(0,\ \max_{b,\,t}\ \bm{r}_b(t)\cdot u\Big),
  \label{eq:g}
\end{equation}
which measures how effectively that direction pushes the system toward failure.
It is zero for directions that never push any line toward its limit.
Such directions never fail to linear order, so the linear prediction rises toward the fraction of directions with $g(u)>0$ rather than toward one; this fraction is one in the grids but 83 to 91\% in the networks of Sec.~\ref{sec:nonlinear}.
A kick fails, to linear order, if $g(u) > 1/B$, so the linear prediction of the failure-probability curve is
\begin{equation}
  \Plin(B) = \text{fraction of directions with } g(u) > 1/B .
  \label{eq:Plin}
\end{equation}

The response vector has a simple meaning.
Its direction, $\bm{r}_b(t)/|\bm{r}_b(t)|$, is the kick that loads line $b$ most effectively at time $t$.
In the language of non-modal stability, this is the optimal perturbation for that line and that target time~\cite{ButlerFarrell1992,Schmid2007Nonmodal}; in power-system analysis, $\bm\ell_b(t)$ is called the trajectory sensitivity of the line flow to the initial state~\cite{HiskensPai2000}.
The length of $\bm{r}_b(t)$ tells how much of the margin a unit kick in that direction consumes, so the longest response vector, $G=\max_{b,t}|\bm{r}_b(t)|$, defines the worst case.
To linear order, no kick smaller than $B_{\rm on}=1/G$ can cause failure, and a kick of that size causes failure only if it points exactly along the longest response vector.
The corresponding worst-case energy is $E_{\rm on}\propto 1/G^2$.

The prediction is cheap.
The response vectors come from one run of the linearized dynamics, and evaluating Eq.~\eqref{eq:g} for thousands of directions is a single matrix product.
For the grids below this takes less than a second.
The nonlinear reference calculation, which locates the failure threshold of every direction by repeated simulation, takes about a minute on ten processor cores.

\subsection{Test on two benchmark grids}

We check the prediction on two standard benchmark grids.
The first is the IEEE 39-bus New England system~\cite{Athay1979,Pai1989,Zimmerman2011}, with 10 generators and 46 lines; a bus is a node of the grid.
The second is the RTS-GMLC system~\cite{Barrows2020}, an updated version of the IEEE Reliability Test System with 73 buses, 120 lines, and synchronous generators at 30 buses.
Both are widely used, with published network data, operating points, and line ratings, the largest power each line may carry.

We model each generator by the swing equation, Newton's equation for its rotor angle, with the published inertias and reactances.
We add speed governors, which adjust a generator's mechanical power when its speed deviates, and treat the loads as fixed impedances.
This reduces each grid to a network of coupled oscillators~\cite{Motter2013,NishikawaMotter2015}.
The benchmark data contain no damping.
We assume a damping coefficient $D=4$ per unit, that is, in units normalized to each machine's rating.
This gives damping ratios of 2 to 10\% to the electromechanical modes, the oscillations in which generators swing against each other; Sec.~\ref{sec:grids} halves and doubles $D$.
A line fails if its apparent power, the magnitude of the complex power it carries, exceeds its rating within 10~s of the kick.
This instantaneous criterion is a proxy.
Real lines tolerate short overloads, so it marks a transient overload that protection systems or operators must handle, not necessarily an outage; Sec.~\ref{sec:duration} tests overloads that must last.
Appendix~\ref{app:grids} gives the details.

\begin{figure*}[t]
  \centering
  \includegraphics[width=0.9\textwidth]{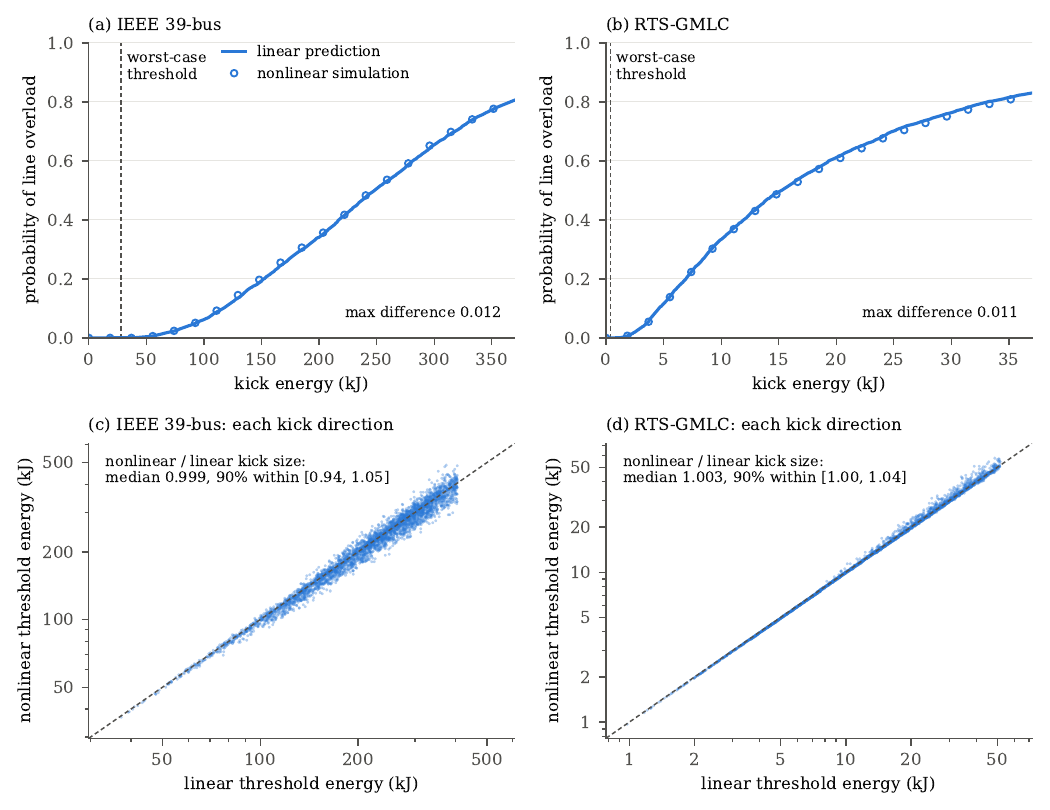}
  \caption{The linear prediction agrees with nonlinear simulations on two benchmark grids.
  (a,b)~Probability that a random frequency kick of given energy overloads a line, from the linear prediction (line) and from nonlinear simulations over 4096 random directions (symbols).
  The dashed line marks the worst-case threshold.
  (c,d)~Failure threshold of each kick direction from nonlinear simulation against the linear prediction.
  Shown are the directions whose linear threshold lies within the simulated range with a 15\% margin (3483 and 3694 of 4096), so that no point is cut off.
  The dashed line is equality.
  Both grids use the model with governors and damping $D=4$ (Appendix~\ref{app:grids}).}
  \label{fig:validation}
\end{figure*}

Figure~\ref{fig:validation} shows the result.
The linear prediction and the nonlinear simulations agree everywhere: the largest difference between the two failure-probability curves is 0.012 for the IEEE grid and 0.011 for RTS-GMLC [Fig.~\ref{fig:validation}(a,b)].
The agreement holds direction by direction, not only on average.
For the smallest kick that overloads a line in a given direction, the ratio of the nonlinear to the linear value has median 0.999 (IEEE) and 1.003 (RTS-GMLC) [Fig.~\ref{fig:validation}(c,d)].
A typical direction is off by 1.8\% and 0.4\%, respectively, and more than 90\% of directions are off by less than 6\%.
We obtained the same agreement, with differences between the curves of at most 0.022, in every variant of the grid models we tried, including other damping values, kicks that also displace the generator angles, and models without governors.
In these grids, lines are overloaded before the generators swing far enough for the nonlinearity of the swing equations to matter.
Section~\ref{sec:nonlinear} discusses a system where this is not so.

Figure~\ref{fig:validation} also shows how little the worst case says about the bulk of the risk.
The worst-case threshold lies far to the left of the rise: the kick energy at which half of all directions fail is 9 times the worst-case energy for the IEEE grid and about 40 times for RTS-GMLC.
The next section explains what sets the rise.

% =============================================================================
\section{The response hull}
\label{sec:hull}

\subsection{A geometric picture}

Collect all the response vectors $\bm{r}_b(t)$, for every line and every time, as points in the space of kick directions, and add the origin.
The smallest convex body that contains these points is the response hull [Fig.~\ref{fig:concept}(a)].
Equation~\eqref{eq:g} has a simple geometric meaning in terms of this body.
The number $g(u)$ is the largest projection of the hull onto the direction $u$, that is, the distance from the origin to the plane perpendicular to $u$ that just touches the hull.
(In geometry, $g$ is called the support function of the hull.)
We call it the reach of the hull in the direction $u$; without a direction, ``the reach'' means its largest value, the hull's largest radius.
The response vector that attains the maximum in Eq.~\eqref{eq:g}, a particular line at a particular moment, is the failure route of the direction $u$.
A kick of size $B$ in direction $u$ therefore causes failure exactly when the hull crosses the plane $\bm{x}\cdot u=1/B$.
The linear failure probability is the fraction of directions in which it does.
For Gaussian kicks, $g$ is the maximum of a Gaussian random process indexed by line and time, and the theory of such processes~\cite{AdlerTaylor2007} supplies the formulas we use below (Appendix~\ref{app:geometry}).

This picture has an immediate consequence: only the shape of the hull matters.
A rotation of the kick coordinates that preserves kick size, such as relabeling the generators, rotates the hull but leaves the distribution of $g(u)$ unchanged.
The risk changes only when the shape of the hull changes, that is, when the dynamics change or when disturbance size is measured differently.
Three features of that shape turn out to control the failure-probability curve.

\subsection{Rule 1: the reach sets where risk begins}

The hull's largest radius is the length $G$ of the longest response vector.
It sets the smallest kick that can cause failure, $B_{\rm on}=1/G$, and it is what a worst-case analysis computes.
Just above this threshold, the dangerous directions form a small cap on the sphere of directions, centered on the longest response vector.
If the response vector kept its direction near its peak, that cap would be the whole story.
But in a dynamical system the optimal direction moves with time.
As the response passes through its peak it sweeps the cap sideways and stretches it into an ellipse, by a factor $\sqrt{1+\Lambda^2}$.
Here the turning number $\Lambda$ is the angle, in radians, through which the response direction turns while the response stays near its peak (Appendix~\ref{app:geometry}).
Each separate peak of the response, for example each swing of a grid, contributes its own stretched cap.
Near onset, the linear failure probability is therefore the worst-case cap, enlarged by turning and repeated for every comparable peak.
We tested this rule on the IEEE grid without governors and with uniform damping, where the largest peak has $\Lambda\approx 0.36$.
It agrees with direct sampling of $2\times10^7$ directions to within 7\% for kicks 1.1 to 1.3 times the worst-case size.
Closer to onset, failures are too rare to sample precisely.

The near-onset regime is narrow, however.
In high dimensions the cap is tiny unless the kick is well above the threshold, so the onset says little about where failures become common, as Fig.~\ref{fig:validation} showed.

\subsection{Rule 2: the mean reach sets where risk is large}

Where failures become common is set by how far the hull reaches in a typical direction.
We call the average of $g(u)$ over all directions the mean reach $\langle g\rangle$; it equals half of what geometers call the mean width of the hull.
Kicks start to fail in large numbers when $1/B$ falls to about $\langle g\rangle$, that is, at kick sizes near $1/\langle g\rangle$.
This is several times the worst-case threshold $1/G$.

The mean reach is where the dynamics enters.
A response that points in one fixed direction makes a thin, needle-like hull: it reaches far along one direction and hardly at all along the others.
A response that turns, or that peaks several times in different directions, fills out the hull and raises its mean reach, even if its longest vector stays the same.
A comparison theorem for Gaussian processes, the Sudakov--Fernique inequality~\cite{Vershynin2018}, makes this precise.
If a change of the dynamics increases every distance between the points that span the hull (the response vectors and the origin), the mean reach cannot decrease (Appendix~\ref{app:geometry}).
A more intuitive measure of the mean reach is the effective number of independent dangerous directions: the number of equally strong, mutually perpendicular response vectors that would produce the same mean reach (Appendix~\ref{app:geometry}).
It is between 2.8 and 8.2 for the grid models we studied.

\subsection{Rule 3: the overlap of failure routes sets how sharply risk rises}

The mean reach locates the rise of the failure-probability curve; its width is set by how much $g(u)$ varies from one direction to another.
A general result on the concentration of measure~\cite{Ledoux2001} guarantees that its standard deviation is at most $G/\sqrt{n-1}$, where $n$ is the number of independent components of a kick.
In the grids it is about two to three times smaller.
The reason is that the variation depends on how the failure routes of neighboring kicks overlap~\cite{Chatterjee2014} (Appendix~\ref{app:geometry}).
If two similar kicks overload the same line at the same moment, their thresholds are correlated; if they overload different lines, they are not.
The more distinct failure routes a system has, the more sharply its risk rises.

\subsection{Consequence: same worst case, shifted risk}

These rules explain the example of Fig.~\ref{fig:concept}.
Suppose the damping or the inertia of a grid changes while the worst case is held fixed.
The reach then stays the same, and so, nearly, does the spread of $g(u)$ (it changes by less than 2\% in our grids), but the mean reach changes.
Moreover, almost every direction gains or loses about the same amount of reach, so the hull thickens or thins nearly uniformly; Sec.~\ref{sec:grids} traces this to the swings of the grid.
Across all twelve grid variants, the correlation coefficient between $g(u)$ before and after the change lies between 0.95 and 0.997.
When the directions are sorted by reach and grouped into bands, each band changes on average by 0.6 to 1.3 times the overall mean change.
Only in the top 1\%, next to the worst case, is the change smaller (0 to 0.6 times the mean), as it must be when the reach is fixed.

\begin{figure*}[t]
  \centering
  \includegraphics[width=\textwidth]{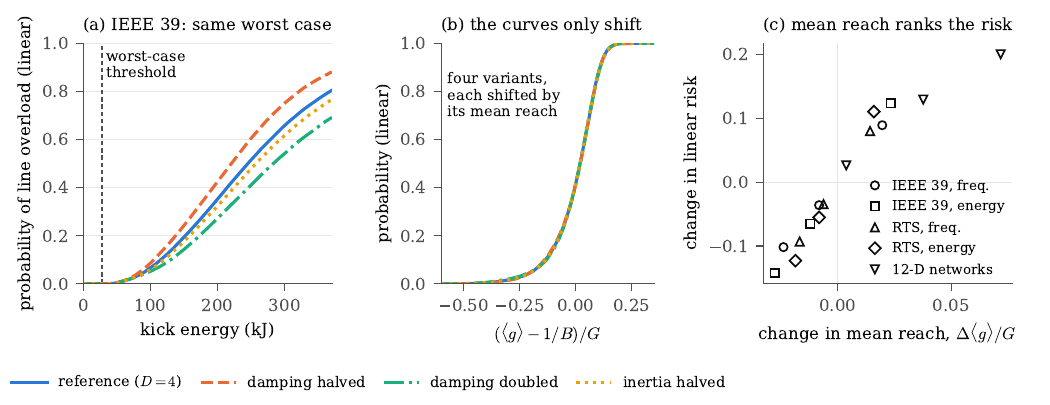}
  \caption{With the worst case fixed, the mean reach of the response hull sets the risk.
  (a)~Linear failure-probability curves of the IEEE 39-bus grid with frequency kicks for the reference model and three variants (damping halved or doubled, inertia halved).
  Line ratings are rescaled so that all four have the same worst-case threshold.
  (b)~The same curves plotted against $(\langle g\rangle-1/B)/G$, that is, shifted by each variant's mean reach $\langle g\rangle$ and scaled by the common reach $G$: they collapse.
  (c)~Change in linear risk, at the kick size where half of all directions fail in the reference model, against the change in mean reach, for all variants of five families: two grids with two kick types each, and the 12-dimensional networks of Sec.~\ref{sec:nonlinear} (relative to network A).}
  \label{fig:shift}
\end{figure*}

The failure-probability curves should then differ almost only by a shift, and Fig.~\ref{fig:shift} confirms this; we call it the shift rule.
In the IEEE grid, four variants with the same worst case have visibly different curves [Fig.~\ref{fig:shift}(a)], but they collapse onto one curve once each is shifted by its own mean reach [Fig.~\ref{fig:shift}(b)].
For both grids and both kick types, the shifted curves agree with the actual ones to within 0.014, while the unshifted curves differ by up to 0.16.
The shape of the curve is not universal (a Gaussian with the right mean and width misses it by up to 0.1), but a change of damping or inertia preserves it.
The mean reach alone then tells which variant is riskier and by how much.

The shift rule is specific to changes of damping and inertia.
The 12-dimensional networks of Sec.~\ref{sec:nonlinear} differ in structure rather than in damping.
Their change of $g(u)$ is far from uniform (correlation coefficients 0.2 to 0.7), and their curves are not shifts of one another.
Even there the mean reach ranks the linear risk correctly, as it does in every family we studied [Fig.~\ref{fig:shift}(c)], although Sec.~\ref{sec:nonlinear} shows that it does not rank their nonlinear risk.

\subsection{What the geometry guarantees}
\label{sec:guarantee}

The shift rule is an observation about our examples, so it is fair to ask how much of the picture is guaranteed.
One statement about risk holds without exception.
If the hull of one system lies inside the hull of another, the first system is at most as risky at every kick size, because its reach is smaller in every direction.
This holds for any distribution of kick directions.
Damping gives an example.
In the idealized swing model of Sec.~\ref{sec:damping}, where damping multiplies every response vector by the envelope $e^{-\gamma t/2}$, stronger damping shrinks every response vector toward the origin. The strongly damped hull then lies inside the weakly damped one, and at fixed line ratings more damping can only lower the linear risk. In the full swing equations this holds only approximately, because damping also slightly changes the oscillation itself.

The mean reach carries no such guarantee.
Two hulls with the same reach can have failure-probability curves that cross: the hull with the larger mean reach is then safer for small kicks and riskier for large ones (Appendix~\ref{app:geometry} gives a two-dimensional example).
That the mean reach ranks the risk in all our families is therefore a finding about these families, not a theorem.
It holds because damping and inertia change the hull nearly uniformly.
Once the worst case is matched, the hulls are no longer nested, and Sec.~\ref{sec:quantile} replaces the ranking by a law that predicts the change of risk at every level.

% =============================================================================
\section{Damping and inertia in power grids}
\label{sec:grids}

\begin{figure*}[t]
  \centering
  \includegraphics[width=\textwidth]{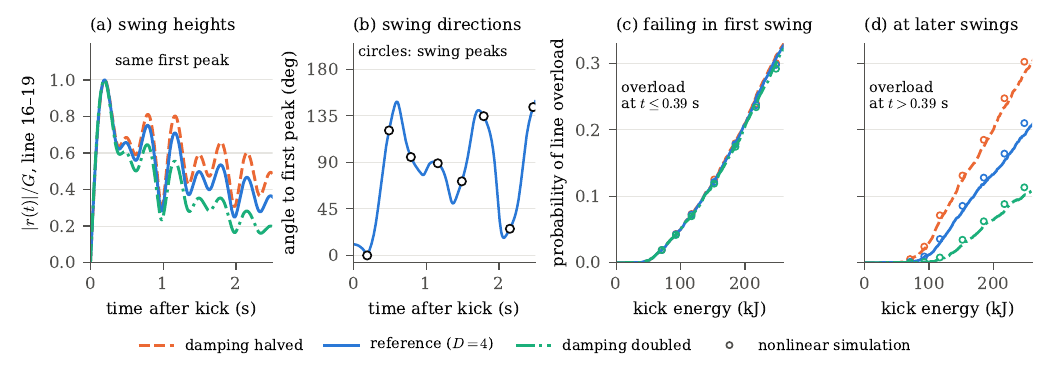}
  \caption{Why damping changes the risk but not the worst case: IEEE 39-bus grid, frequency kicks, three damping values with the same worst case.
  (a)~Length of the response vector of the most heavily loaded line (buses 16--19), relative to the worst case.
  The first peak, which sets the worst case, is the same for all three damping values; later swings are smaller with stronger damping.
  (b)~Angle between this line's response direction and its direction at the first peak (reference model); circles mark the swing peaks.
  Later swings point in directions far from the first, so each adds a new dangerous region of kick directions.
  (c)~Probability that a kick first overloads a line during the first swing ($t\le t_1=0.39$~s) and (d)~at a later swing.
  Lines: linear prediction; symbols: nonlinear simulations over 4096 directions.
  First-swing failures are the same for all three damping values; damping acts on later-swing failures only.}
  \label{fig:mechanism}
\end{figure*}

\begin{table*}[t]
  \caption{Same worst case, different risk.
  Probability that a random kick of energy $E$ overloads a line, from nonlinear simulations over 4096 directions, for the reference model ($D=4$) and three variants with the same worst-case threshold $E_{\rm on}$: damping halved ($D/2$), damping doubled ($2D$), and inertia halved ($H/2$).
  Kicks change the generator frequencies (freq.) or both angles and frequencies (angle+freq.).
  The linear predictions agree with these values to within 0.009.
  The last column is the ratio of the risks for $D/2$ and $2D$.
  All variants use the same 4096 directions, so differences between columns are much less noisy than the $\pm0.01$ precision of each entry.}
  \label{tab:matched}
  \begin{ruledtabular}
  \begin{tabular}{llccccccc}
  Grid & Kicks & $E_{\rm on}$ (kJ) & $E$ (kJ) & Reference & $D/2$ & $2D$ & $H/2$ & Ratio \\
  \hline
  IEEE 39  & freq.       & 28   & 135 & 0.160 & 0.199 & 0.115 & 0.147 & 1.7 \\
           &             &      & 220 & 0.410 & 0.499 & 0.327 & 0.383 & 1.5 \\
  IEEE 39  & angle+freq. & 21   & 120 & 0.122 & 0.173 & 0.081 & 0.104 & 2.1 \\
           &             &      & 200 & 0.363 & 0.486 & 0.249 & 0.316 & 2.0 \\
  RTS-GMLC & freq.       & 0.37 & 8   & 0.247 & 0.305 & 0.187 & 0.224 & 1.6 \\
           &             &      & 15  & 0.490 & 0.567 & 0.397 & 0.458 & 1.4 \\
  RTS-GMLC & angle+freq. & 0.30 & 6   & 0.164 & 0.225 & 0.115 & 0.138 & 2.0 \\
           &             &      & 11  & 0.404 & 0.511 & 0.298 & 0.358 & 1.7 \\
  \end{tabular}
  \end{ruledtabular}
\end{table*}

This section looks inside the two grids to find where the shift of the risk comes from.
We first show why their responses turn, then locate the failures that damping changes, explain why, and turn the explanation into a law for the whole failure-probability curve.
Finally, we contrast damping with inertia and check the conclusions against a stricter failure criterion.

\subsection{Why the response of a grid turns}
\label{sec:turning}

After a kick, the generators swing against each other.
Each swing loads the lines anew, and each is a different mixture of the grid's oscillation modes, so the kick direction that loads a given line most effectively changes from one swing to the next.
Figure~\ref{fig:mechanism} shows this for the most heavily loaded line of the IEEE grid, between buses 16 and 19.
Its response peaks 0.19~s after the kick and again roughly every 0.3 to 0.4~s [Fig.~\ref{fig:mechanism}(a)].
At the next five peaks it points at angles of $72^\circ$ to $135^\circ$ from its direction at the first peak [Fig.~\ref{fig:mechanism}(b)], so a kick that misses the first peak can still overload the line at a later swing.
Other lines add peaks of their own: in the reference model, six further lines reach more than half of the worst-case response.
Every swing thus adds a new vertex to the response hull, pointing in a new direction.

Damping acts on these vertices unevenly.
The first peak comes so soon after the kick that damping has little time to act, so it is nearly the same for weak and strong damping, and it is this peak that sets the worst case.
The later peaks are different.
With the damping halved, the next three peaks of line 16--19 reach 69, 81, and 80\% of the worst-case response; with the damping doubled, they reach only 61, 65, and 56\%.
Damping therefore trims the later vertices of the hull and leaves its longest one in place.
It lowers the mean reach but not the reach: the effective number of independent dangerous directions falls from 4.9 with halved damping to 3.9 with doubled damping.
Judged by their least-damped oscillation mode, the halved and doubled variants represent a poorly damped and a well-damped grid (Appendix~\ref{app:grids}).

\subsection{Damping changes the risk at a fixed worst case}

If damping trims only the later vertices of the hull, grids that differ in damping but share a worst case should differ in risk.
Table~\ref{tab:matched} tests this in both grids and for two kinds of kicks: frequency kicks, and kicks that also displace the generator angles.

For frequency kicks the first peak, and with it the worst case, still changes slightly with damping.
We therefore rescale the line ratings of each variant by a common factor, of about 1\%, so that its worst-case threshold equals that of the reference model.
The rescaling works slightly against the effect we report, and it is a convenience rather than a requirement: with the original ratings the results are similar (Appendix~\ref{app:grids}).
The angle-displacing kicks need no rescaling, because their worst kick is a pure angle displacement that overloads a line at the instant of the kick, before damping or inertia can act.

The results are consistent across grids and kick types.
Over the whole range where 10 to 50\% of kicks fail, halving the damping instead of doubling it raises the risk by a factor of 1.4 to 2.2; at the eight energies of Table~\ref{tab:matched} the factor is 1.4 to 2.1, and the linear prediction matches the simulations to within 0.009.
Near the worst-case threshold, by contrast, the variants nearly coincide: up to about 2.5 times the worst-case energy, fewer than 3\% of kicks fail, and the variants differ by less than 0.005.
The same pattern holds for kicks at a single, randomly chosen generator, the disturbances of basin-stability studies~\cite{Menck2013BasinStability} (Appendix~\ref{app:grids}).
Halving the inertia at fixed kick energy slightly lowers the risk; Sec.~\ref{sec:inertia} returns to inertia with a fairer comparison.

\subsection{First-swing and later-swing failures}
\label{sec:swings}

Section~\ref{sec:turning} suggests that damping changes the risk through the later swings.
To test this, we record for each kick when a line is first overloaded, and we call the failure a first-swing failure if this happens before the first swing ends.
We take the end of the first swing to be the first minimum of the worst line's response after its peak: $t_1=0.39$~s in the IEEE grid and 0.44~s in RTS-GMLC.

Figure~\ref{fig:mechanism}(c,d) shows the result for frequency kicks in the IEEE grid.
First-swing failures are the same for all three damping values.
Later-swing failures are not: at 185~kJ, where 30\% of kicks fail in the reference model, they occur for 18.5\% of kicks with the damping halved and for 6.2\% with it doubled.
The nonlinear simulations show the same split, and the linear prediction agrees with them to within 0.02 for both parts.

RTS-GMLC behaves in the same way.
For frequency kicks in both grids, and at every kick energy where 2 to 50\% of kicks fail, at least 94\% of the difference between halved and doubled damping comes from later-swing failures.
The split does not hinge on exactly where the first swing is taken to end (Appendix~\ref{app:grids}).
For kicks that also displace the angles, the later swings carry 57 to 82\% of the difference, because damping already acts on the speed changes that such kicks build up during the first swing.
Later-swing failures are not a small correction.
With frequency kicks they make up 33 to 55\% of all failures in the reference models, at energies where 10 to 50\% of kicks fail.

The hull explains this split and gives a second threshold.
The first-swing response vectors include the longest one and are fixed before damping has had time to act.
The later ones reach at most $\rho G$, where $\rho<1$ is the length of the longest later-swing response vector relative to the worst case.
To linear order, no kick smaller than $B_2=B_{\rm on}/\rho$ can fail at a later swing, so below the kick energy $E_2=E_{\rm on}/\rho^2$ damping can change the risk only through the first swing, which it hardly touches.
In the IEEE grid, $\rho=0.81$, 0.75, and 0.65 for halved, reference, and doubled damping, so $E_2$ is 1.5, 1.8, and 2.4 times the worst-case energy; in all our grid families it lies between 1.4 and 2.4 times the worst-case energy.
Above $E_2$ the variants separate quickly: at 2.5 and 3 times the worst-case energy, where only 2 to 5\% of kicks fail, halved damping is already 1.2 and 1.5 times riskier than doubled damping (linear prediction, $4\times10^4$ directions).

\subsection{Damping discounts late failure routes}
\label{sec:damping}

A short calculation shows why damping acts on late failures.
Suppose every generator has the same damping rate $\gamma$ and there are no governors.
The linearized swing equations then give $\bm r_b(t;\gamma)=e^{-\gamma t/2}\,\bm r_b(t;0)$, up to corrections of relative order $(\gamma/\omega)^2\approx0.01$, where $\omega$ is the frequency of the swings (Appendix~\ref{app:geometry}).
Every response vector is discounted by a factor that depends only on its time.
The reach $g(u)$ of a direction is attained by its failure route at a time $t_*(u)$, which we call the failure time of that direction.
A small change of the damping rate therefore changes the reach by
\begin{equation}
  \frac{\partial \ln g(u)}{\partial\gamma}=-\frac{t_*(u)}{2}.
  \label{eq:tlaw}
\end{equation}
Damping discounts each failure route according to how late it occurs.
The worst case, whose route peaks at 0.19~s, changes by only a few percent when the damping is doubled, while routes through later swings shrink much more.

Equation~\eqref{eq:tlaw} holds closely even in the full models, where the damping rates differ between machines and the governors add damping of their own.
Across directions, the exact derivative $\partial\ln g/\partial\ln D$ and $-\bar\gamma\, t_*/2$, with $\bar\gamma$ the mean damping rate, have correlation coefficient $-0.99$ in the IEEE grid and $-0.98$ in RTS-GMLC (Appendix~\ref{app:geometry}).

In practice, this means that the effect of damping on the risk can be estimated from the failure times of the reference model, without simulating any variant.
Averaged over directions, Eq.~\eqref{eq:tlaw} gives the change of the mean reach, $\partial\langle g\rangle/\partial\gamma=-\langle t_*\,g\rangle/2$.
For the matched variants of Table~\ref{tab:matched}, whose ratings are rescaled to restore the worst case, the estimate is $g(u)\to g(u)\,\exp\{-\Delta\gamma\,[t_*(u)-t_G]/2\}$, with $t_G$ the time of the worst case.
It reproduces their linear failure-probability curves to within 0.012 to 0.059, against raw differences of 0.08 to 0.17.

\subsection{How damping moves the whole curve}
\label{sec:quantile}

So far we have followed one direction at a time.
To see how the whole failure-probability curve moves, we need one more step, and this step is exact.
Let $E_p$ be the kick energy at which a fraction $p$ of all kicks fails, with the line ratings held fixed, and let $E_{\rm on}$ be the worst-case energy at the same damping.
When the damping changes, $E_p$ can move only through the directions that sit exactly on the failure threshold at that energy, because these are the ones that switch between failing and surviving.
The threshold moves by their average change, an exact identity known from the sensitivity analysis of quantiles~\cite{Hong2009}.

To use this identity, define the damping susceptibility of a direction, $s(u)=-\partial\ln g(u)/\partial\ln D$: the relative amount by which its reach shrinks when the damping of every machine grows by a common factor.
Let $s_G$ be the same quantity for the worst case.
Then (Appendix~\ref{app:geometry})
\begin{equation}
  \frac{d\ln(E_p/E_{\rm on})}{d\ln D}=2\,\big(\langle s\rangle_p-s_G\big),
  \label{eq:qexact}
\end{equation}
where $\langle s\rangle_p$ is the mean susceptibility of the kicks on the threshold.
Within the linear response, and under the mild smoothness conditions stated in Appendix~\ref{app:geometry}, this susceptibility law is exact.
It holds for any distribution of damping among the machines, with or without governors.
Computing $s(u)$ for all directions takes only two additional linear simulations, with the damping slightly raised and lowered.

The discount of Sec.~\ref{sec:damping} makes the susceptibility law transparent.
If every machine had the same damping rate $\gamma$, the susceptibility of a direction would be $s(u)=\gamma\,t_*(u)/2$, and Eq.~\eqref{eq:qexact} would become
\begin{equation}
  \frac{d\ln(E_p/E_{\rm on})}{d\ln D}=\bar\gamma\,\big(\langle t_*\rangle_p-t_G\big),
  \label{eq:qlaw}
\end{equation}
where $\langle t_*\rangle_p$ is the mean failure time of the kicks on the threshold and $t_G$ is the time of the worst case.
We call Eq.~\eqref{eq:qlaw} the time law.
In the full models, where the machines have different damping rates, we take $\bar\gamma$ to be their arithmetic mean (Appendix~\ref{app:geometry}).
This replacement is an approximation, not part of Eq.~\eqref{eq:qexact}.

\begin{figure}[t]
  \centering
  \includegraphics[width=\columnwidth]{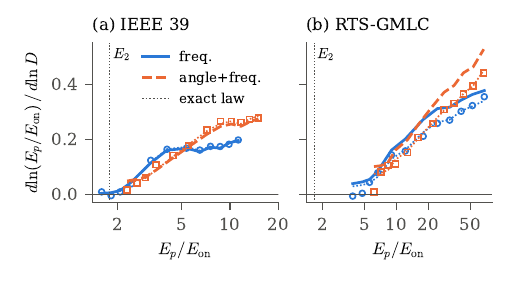}
  \caption{How damping moves the failure-probability curve.
  Relative change of $E_p/E_{\rm on}$, the kick energy at which a fraction $p$ of kicks fails in units of the worst-case energy at the same damping, per relative change of the damping of all machines, plotted against $E_p/E_{\rm on}$ for levels $p$ from 0.2 to 70\%.
  Thick lines: time law, Eq.~\eqref{eq:qlaw}, from the reference model alone; thin dotted lines: exact law, Eq.~\eqref{eq:qexact}; symbols: finite differences with all damping scaled by $1\pm0.1$ at fixed line ratings ($4\times10^4$ directions).
  Frequency kicks (solid, circles) and kicks that also displace the angles (dashed, squares).
  The dotted line marks the second threshold $E_2$ for frequency kicks.}
  \label{fig:quantile}
\end{figure}

In words, damping moves the curve at level $p$ in proportion to how much later than the worst case the kicks at that level fail.
Near onset, the kicks on the threshold fail through the worst-case route itself, so $\langle t_*\rangle_p\approx t_G$ and the curve stays put.
Where failures are common, the kicks on the threshold fail at later swings, 0.2 to 0.9~s after the worst case, and stronger damping pushes the curve to larger energies.
The sensitivity therefore switches on above the second threshold $E_2$ of Sec.~\ref{sec:swings}, where later swings begin to fail (Fig.~\ref{fig:quantile}).

Figure~\ref{fig:quantile} tests both laws against finite differences, obtained by recomputing the curve with all damping slightly raised and lowered.
The susceptibility law, Eq.~\eqref{eq:qexact}, agrees with them to within 0.015 in both grids and for both kinds of kicks, at every level where 5 to 70\% of kicks fail.
The time law, Eq.~\eqref{eq:qlaw}, agrees to within 0.010 (IEEE) and 0.012 (RTS-GMLC) in the idealized model with uniform damping and no governors, and to within 0.018 in the full IEEE model.
In the full RTS-GMLC model it overestimates the sensitivity by up to 30\%.
Because the susceptibility law still holds there, this gap measures the common-damping approximation alone: the damping rates differ more between machines, and the governors add damping that does not change with $D$.

The sensitivities $d\ln(E_p/E_{\rm on})/d\ln D$ are modest, 0.1 to 0.5, but the curves are steep.
From halved to doubled damping, a factor of four, the energy at which half of all kicks fail rises by 27\% relative to the worst case in the IEEE grid and by 53\% in RTS-GMLC.
Shifts of this size produce risk ratios like those of Table~\ref{tab:matched}.

\subsection{Inertia rescales the hull, damping reshapes it}
\label{sec:inertia}

Inertia, the other generator property that we vary, acts differently from damping.
To compare grids with different inertia fairly, we kick them with the same power imbalance rather than the same energy.
A short imbalance $J_i$ at generator $i$ changes its speed by $\omega_s J_i/(2H_i)$, so a lighter grid receives a larger kick.
We draw the imbalances uniformly in direction and measure their size $|J|$ in MW\,s (Appendix~\ref{app:grids}).

Without governors, the linearized swing equations obey a scaling law for such kicks (Appendix~\ref{app:geometry}).
Multiply every inertia by $\lambda$ and every damping rate by $\lambda^{-1/2}$.
The response to the same imbalance is then the old response, compressed in time by $\sqrt\lambda$ and amplified by $\lambda^{-1/2}$.
The time compression does not matter here: in our grids every failure route peaks well inside the 10~s window, and extending the window to 20~s changes the reach of no direction.
The whole hull is therefore multiplied by $\lambda^{-1/2}$, and the failure-probability curve is simply rescaled, $P_\lambda(J)=P_1(J/\sqrt\lambda)$.
The worst case changes by the same factor, so in this limit it predicts the change of risk exactly.

\begin{figure*}[t]
  \centering
  \includegraphics[width=\textwidth]{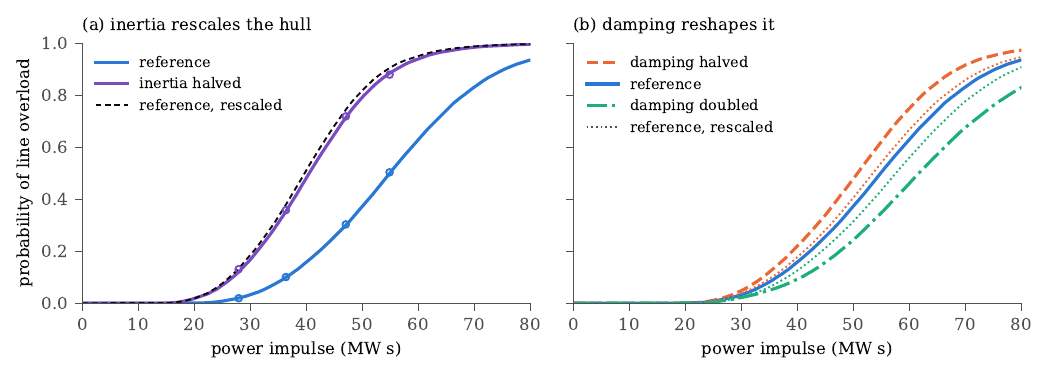}
  \caption{Inertia rescales the response hull, damping reshapes it: IEEE 39-bus grid, kicks that are short power imbalances of a given size, no rescaling of line ratings.
  (a)~Halving the inertia.
  The low-inertia curve is the reference curve rescaled by the ratio of the worst cases (dashed).
  (b)~Halving or doubling the damping.
  Rescaling the reference curve by the ratio of the worst cases (dotted) misses most of the change.
  Lines: linear prediction; symbols in (a): nonlinear simulations over 4096 directions.}
  \label{fig:inertia}
\end{figure*}

Realistic changes stay close to this limit.
Halving the inertia of the generators with the same per-unit damping doubles the damping rate rather than multiplying it by $\sqrt2$, and the governors keep their time constants.
Still, the reach grows by a factor of 1.38 in both grids, and the mean reach almost as much, by 1.36 (IEEE) and 1.34 (RTS-GMLC).
Rescaling the reference curve by the ratio of the worst cases therefore reproduces the low-inertia curve to within 0.033 (IEEE) and 0.041 (RTS-GMLC), although the two curves differ by up to 0.42 and 0.28 [Fig.~\ref{fig:inertia}(a)].
For the reference and low-inertia grids, the linear prediction agrees with the nonlinear simulations to within 0.02.
Unlike damping, inertia also acts on the first swing.
At an imbalance of 36~MW\,s, for example, halving the inertia raises the risk of the IEEE grid from 0.10 to 0.36 in the nonlinear simulations, and most of this increase, 0.19, comes from first-swing failures.

Damping behaves in the opposite way [Fig.~\ref{fig:inertia}(b)].
At fixed power imbalance, halving or doubling it changes the worst case by less than 5\% but the mean reach by 8 to 14\%.
Rescaling by the ratio of the worst cases then misses most of the change, leaving errors of 0.085 to 0.11 against raw differences of 0.11 to 0.17, whereas rescaling by the ratio of the mean reaches works to within 0.031.

In short, inertia sets the scale of the response hull, and damping sets its shape.
A worst-case analysis sees a loss of inertia but not a loss of damping.

\subsection{Overloads that must last}
\label{sec:duration}

So far a line fails as soon as its flow exceeds the rating.
Protection systems and operators, however, act only on overloads that persist.
We therefore repeated the matched comparison with frequency kicks, counting a failure only if a line stays overloaded for a time $\tau$ of 0.1 or 0.2~s without interruption (Appendix~\ref{app:grids}).
The linear prediction carries over.
The reach of a direction becomes the largest overload that some line sustains for the time $\tau$, which still grows in proportion to the kick, although it is no longer the support function of a convex body.

Requiring the overload to last moves the risk to larger kicks.
In the IEEE grid, half of all kicks fail at 248~kJ when any overload counts, at 300~kJ for $\tau=0.1$~s, and at 538~kJ for $\tau=0.2$~s.
The conclusions do not change.
Where 10 to 50\% of kicks fail, halving instead of doubling the damping raises the risk by a factor of 1.4 to 1.8 in both grids, and at least 89\% of the difference comes from failures after the first swing.
The linear prediction stays within 0.021 of the simulations, except in the IEEE grid with $\tau=0.2$~s, where it underestimates the risk by up to 0.059.

% =============================================================================
\section{When the linear picture fails}
\label{sec:nonlinear}

In the grids of Secs.~\ref{sec:linear}--\ref{sec:grids}, the nonlinear thresholds stay within a few percent of the linear ones.
To see what happens when the nonlinearity is strong, we use small model networks with 12 variables that obey $\dot x = Ax + \beta\,x\circ x$, where $x\circ x$ squares every variable separately (Appendix~\ref{app:networks}).
The matrix $A$ is stable and feed-forward: it amplifies signals transiently on their way to an output variable $x_1$, and the network fails when $x_1$ reaches 1.
Network A has a single amplifying chain.
Network B has three identical chains, with weaker couplings chosen so that A and B have exactly the same worst case.

\begin{figure*}[t]
  \centering
  \includegraphics[width=\textwidth]{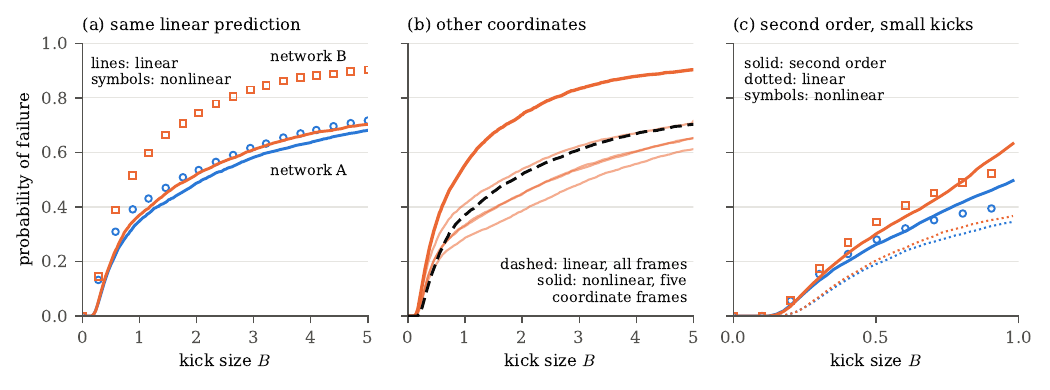}
  \caption{When the linear picture fails: 12-dimensional networks with a strong nonlinearity ($\beta=0.35$).
  (a)~Networks A and B have the same worst case and nearly the same response hull, so their linear predictions (lines) almost coincide; the nonlinear simulations (symbols) differ.
  (b)~Networks that share B's linear dynamics but whose quadratic term acts on rotated variables: one linear prediction (dashed), very different nonlinear risks (solid; the thick line is network B itself).
  (c)~For small kicks, a second-order response calculation (solid) captures most of the nonlinear difference between A and B that the linear prediction (dotted) misses; beyond $B\approx0.8$ it overshoots.}
  \label{fig:nonlinear}
\end{figure*}

Because B's three chains act in unison, its hull has almost the same shape as A's, and their linear failure-probability curves differ by at most 0.04 [Fig.~\ref{fig:nonlinear}(a)].
With a strong nonlinearity, $\beta=0.35$, network B is nevertheless up to 0.22 more likely to fail than A.
The squared terms of B's nine chain variables all feed the output with the same sign.
They push B toward failure even for kicks that the linear response would never carry there: of the 4096 kick directions, 388 cannot cause failure at any size in the linear prediction yet still cause failure in B, against one in A.

This effect depends on which variables the quadratic terms couple, and the linear response cannot see it.
To show this, we take B's linear dynamics, write it in rotated variables, and square the new variables instead of the old ones.
The result is a family of different nonlinear systems that share one linearization.
All of them have B's hull and B's linear prediction, but their nonlinear failure probability, averaged over kick sizes from 0.2 to 5, ranges from 0.43 to 0.73 [Fig.~\ref{fig:nonlinear}(b)].
When the rotation merges B's three chains into one, the risk equals that of A (0.55 for both).

The quadratic coupling enters at second order.
Expanding the response as $x\approx B\,x^{(1)}+B^2x^{(2)}$, where $x^{(1)}$ obeys the linear dynamics and $x^{(2)}$ is driven by $\beta\,x^{(1)}\circ x^{(1)}$, recovers most of what the linear picture misses for small kicks [Fig.~\ref{fig:nonlinear}(c)].
At kick sizes $B\le 0.75$ the second-order prediction reproduces 60\% of the difference between B and A, against 20\% for the linear one.
It cuts the largest error from 0.09 to 0.03 for network A and from 0.16 to 0.04 for network B.
Beyond $B\approx0.8$ it overshoots, and higher orders would be needed.
When the nonlinearity is weak, $\beta=0.05$, the geometric picture is restored (Appendix~\ref{app:networks}).

These examples give a practical rule.
The linear prediction can be trusted when the nonlinear thresholds stay close to the linear ones, which a few nonlinear simulations can check.
In the grids, 90\% of directions have nonlinear thresholds within 6\% of the linear ones (Fig.~\ref{fig:validation}).
In network B at $\beta=0.35$ they do not, and there the structure of the nonlinear coupling becomes part of the answer.

% =============================================================================
\section{Discussion}
\label{sec:discussion}

\subsection{What the worst case leaves out}

Worst-case methods compute the reach of the response hull.
The optimal perturbation~\cite{ButlerFarrell1992,Schmid2007Nonmodal}, the minimal seed~\cite{PringleKerswell2010MinimalSeed,Kerswell2018}, the conditional nonlinear optimal perturbation~\cite{MuDuanWang2003CNOP}, the most probable failure point of structural reliability~\cite{HasoferLind1974FORM,DitlevsenMadsen1996Reliability}, and the minimal desynchronizing disturbance of a power grid~\cite{Eaves2025} all identify the single direction in which the system is closest to failure.
That direction sets where risk begins, not where it becomes large.
In a power grid it belongs to the first swing, whereas damping acts mainly through the later ones.
The missing information is not hard to obtain.
Optimal-perturbation calculations already compute the best kick direction for every target time; the response hull is simply the collection of all of them, which a worst-case analysis reduces to its largest member.
Keeping the whole collection costs little extra.

The worst case remains a useful guide to where to act.
When we add damping to one generator at a time, the generator that lowers the worst case most also lowers the risk most, in both grids.
What the worst case misses is how large the change of risk is: for that generator in the IEEE grid, doubling its damping raises the worst-case energy by only 6\%, while the fraction of failing kicks at the energy where 30\% failed before drops to 25\%.

\subsection{Relation to other work}

The probability we compute is familiar to reliability engineers, who call it a fragility curve when it is plotted against the intensity of the load~\cite{Porter2007}.
In their terms, our linear prediction is a first-order reliability calculation (one that linearizes the failure condition) for a failure condition that changes in time, and the rate at which the boundary is crossed follows from Rice's formula~\cite{Rice1944,AndrieuRenaud2004,Sudret2008}.
For a fixed kick size, the safe directions form a polyhedron, the setting of classical work on polyhedral safe sets~\cite{Ditlevsen1983,HagenTvedt1991}.
In probability theory, the same quantity is the excursion probability of a Gaussian random process, for which tube formulas, comparison theorems, and concentration inequalities are available~\cite{AdlerTaylor2007,Vershynin2018}.
In power systems, chance-constrained optimal power flow~\cite{Bienstock2014}, which schedules generators so that line limits hold with a prescribed probability, uses the same kind of probability, but for steady operating points rather than transients.
What the present work adds is the dynamical content: which features of a transient response set the risk, how damping and inertia change them, and how well the resulting predictions hold up against nonlinear simulations.

The same probability can be estimated by sampling nonlinear simulations, as in basin stability, survivability, and probabilistic studies of transition~\cite{Menck2013BasinStability,Hellmann2016Survivability,Pershin2020}, or by importance sampling of grid contingencies, the outages of lines or generators~\cite{Almada2025}.
Sampling makes no approximation and remains necessary for very large kicks or strong nonlinearity.
The linear prediction is cheaper by orders of magnitude, explains why one system is riskier than another, and can be checked with a few nonlinear runs.
Linear-response studies of power grids~\cite{Manik2017,Tyloo2018} characterize the typical size of the response rather than the probability of crossing a limit.
In the language of response theory~\cite{Ruelle2009}, our prediction uses the first-order response of the quantity that fails, and Sec.~\ref{sec:nonlinear} shows that the structure of the nonlinear coupling enters at second order.

\subsection{A practical recipe}

The method has four steps.
(1)~Linearize the system around its operating state, and compute the response vector of every failure boundary over the time window of interest, Eq.~\eqref{eq:r}.
(2)~Draw random kick directions from the disturbance ensemble of interest, compute $g(u)$ for each, Eq.~\eqref{eq:g}, and read off the failure-probability curve, Eq.~\eqref{eq:Plin}.
(3)~Report three kick sizes: the worst-case threshold $1/G$; the threshold $1/(\rho G)$, above which later swings can cause failure; and the typical failure size $1/\langle g\rangle$.
(4)~Check the linear prediction near the typical failure size with a few nonlinear simulations.
To compare two designs or operating states with the same worst case, compare their mean reach; to estimate the effect of a change of damping, use Eq.~\eqref{eq:qexact}, or its time form, Eq.~\eqref{eq:qlaw}, for a first look.

\subsection{Limitations}

Our statements are of three kinds.
Within the linear response, some are exact: the support-function form of the prediction, its invariance under rotations of the kick, the ordering of risk by hull inclusion, the susceptibility law, Eq.~\eqref{eq:qexact}, and the scaling law for inertia.
Others are approximations with a known error: the onset rule near the threshold, and the damping discount $e^{-\gamma t/2}$ on which the time laws, Eqs.~\eqref{eq:tlaw} and \eqref{eq:qlaw}, rest; with unequal damping rates and governors, these also replace the rates by a common mean.
The shift rule and the ranking by mean reach are empirical: they hold for every change of damping and inertia we tried, but no theorem guarantees them (Sec.~\ref{sec:guarantee}).

The probabilities refer to a chosen ensemble of kicks.
We tested four (frequency kicks, kicks that also displace the angles, kicks at a single generator, and short power imbalances) and reached the same conclusions.
Other ensembles, such as correlated fluctuations of renewable generation, weight the directions differently and would need their own check.

The grid models are deliberately simple: classical generators without voltage control, fixed-impedance loads, and damping values that the benchmarks do not provide.
Failure means that a line rating is exceeded, at once or for 0.1 to 0.2~s; thermal limits act over minutes and call for a different model~\cite{Almada2025}.
The split into first-swing and later-swing failures depends on where the first swing is taken to end.
Moving this point by 20\% changes little, but grids with several comparably stressed lines may need a definition line by line.
Each nonlinear curve rests on 4096 directions, which limits its precision to about $\pm0.01$.
Differences are more precise, because all variants use the same directions: the 95\% confidence half-width is below 0.007 for the difference between the nonlinear and linear curves and 0.006 to 0.014 for the difference between halved and doubled damping (Appendix~\ref{app:grids}).

Finally, the linear picture needs the nonlinearity to be weak at the failure threshold.
When it is strong, the linear prediction fails, and the second-order correction helps only for small kicks.
Failures that are strongly nonlinear in themselves, such as subcritical transition to turbulence in shear flows~\cite{PringleKerswell2010MinimalSeed,Kerswell2018}, lie outside the picture altogether.

\subsection{Conclusion}

Worst-case analysis answers the question of how small a disturbance can cause failure.
For random disturbances the more useful question is how likely failure is.
When the nonlinearity is weak at the failure threshold, as in the power grids studied here, the answer is written in the geometry of the transient response.
The reach of the response hull marks where risk begins, and its mean reach marks where risk becomes large.
In a power grid the two belong to different parts of the dynamics.
The worst case is decided in the first swing, while damping changes the risk through later swings that the worst case does not see; inertia, in contrast, rescales every swing and changes the worst case with the risk.
How far damping moves the risk at a given level is set, approximately, by how much later than the worst case the disturbances at that level fail.
Two grids with the same worst case can therefore differ by a factor of about two in risk, and a single linear calculation tells them apart and explains why.

\begin{acknowledgments}
The author thanks Prof. Dr. Pranut Potiyaraj for encouragement.
No external funding was received for this research.
\end{acknowledgments}

\section*{Data Availability Statement}
The code and saved numerical results that support the findings of this study are openly available in Zenodo at \url{https://doi.org/10.5281/zenodo.23860959}~\cite{limkumnerd_zenodo_archive}.
The benchmark grid data are not redistributed; the archive contains a script that downloads them from their original sources~\cite{Zimmerman2011,Barrows2020} at fixed versions.

\appendix

\section{Grid models and simulations}
\label{app:grids}

\paragraph*{Data.}
For the IEEE 39-bus system we take the network, the solved operating point, and the line ratings from the MATPOWER case file~\cite{Zimmerman2011}.
The generator inertia constants $H_i$ and transient reactances $x'_{d,i}$ are the standard values of Refs.~\onlinecite{Athay1979,Pai1989}.
The generator ratings, used only to scale the governors, are those of the DIgSILENT implementation of the system, as tabulated in Ref.~\onlinecite{Bizzarri2024}.
For RTS-GMLC we use the published bus, branch, and generator tables~\cite{Barrows2020}.
Synchronous units at the same bus are combined into one machine, and solar and wind units, which have no inertia, are treated as negative loads.
The published bus voltages belong to a different dispatch, so we re-solve the AC power flow before building the model.

\paragraph*{Model.}
Each generator is a voltage source of fixed magnitude $|E_i|$ behind its transient reactance, and each load is a fixed impedance.
Eliminating the load buses leaves a network of generators coupled through an effective admittance matrix $Y$~\cite{Motter2013,NishikawaMotter2015}.
With generator angles $\delta_i$ and speed deviations $\omega_i$ (rad/s), the dynamics read
\begin{align}
  \dot\delta_i &= \omega_i, \nonumber\\
  \frac{2H_i}{\omega_s}\,\dot\omega_i &= P_{m,i} + p_i - P_{e,i}(\delta) - \frac{2H_i}{\omega_s}\gamma_i\,\omega_i, \label{eq:swing}\\
  T_g\,\dot p_i &= -p_i - \frac{S_i}{R\,S_{\rm base}}\,\frac{\omega_i}{\omega_s}, \nonumber
\end{align}
where $\omega_s=2\pi\times 60$~rad/s, $P_{e,i}={\rm Re}\,[E_i\sum_j (Y_{ij}E_j)^*]$ is the electrical power, $p_i$ is the governor action with droop $R=0.05$ and time constant $T_g=0.5$~s, and $S_i$ is the machine rating.
The damping rate is $\gamma_i=D/(2h_i)$, where $h_i=H_i S_{\rm base}/S_i$ is the inertia constant on the machine's own rating.
We use $D=4$ unless stated otherwise.
The least-damped electromechanical mode then has a damping ratio of 3.0\% (IEEE) and 2.4\% (RTS-GMLC); halving $D$ lowers these to 1.8 and 1.4\%, and doubling it raises them to 5.6 and 4.5\%.
Because the bus voltages are linear in the generator voltages, the apparent power at both ends of every line follows directly from the generator angles, and a line fails when either end exceeds its rating.

\paragraph*{Kicks.}
A frequency kick sets the speed deviations at time zero, with the angles at their equilibrium values.
Its kinetic energy is $E_{\rm kick}=\sum_i H_i S_{\rm base}(\omega_i/\omega_s)^2$, the rotational energy carried by the speed deviations.
We choose $\omega_i(0)=B\,u_i\sqrt{\omega_s/H_i}$ with $u$ a random unit vector, so that all directions with the same energy are equally likely and $E_{\rm kick}=B^2S_{\rm base}/\omega_s=B^2\times265$~kJ.
In a variant, kicks also displace the angles, with the energy measured by the quadratic form of the transient energy function~\cite{Pai1989}.
Single-generator kicks change the angle and speed of one randomly chosen generator $i$.
Their energy is $H_iS_{\rm base}(\omega_i/\omega_s)^2+\tfrac12 K_{ii}\delta_i^2\,S_{\rm base}/\omega_s$, where $K_{ii}$ is the generator's synchronizing coefficient, and the direction in this two-dimensional plane is uniformly distributed.
Power-impulse kicks (Sec.~\ref{sec:inertia}) set $\omega_i(0)=\omega_s J_i/(2H_i)$, with the impulse vector $J$ (per unit of $S_{\rm base}$ times seconds, so that 1 corresponds to 100~MW\,s) uniformly distributed in direction; its size is $|J|$, and the line ratings are not rescaled.

\paragraph*{Matched worst case.}
For the frequency kicks of Table~\ref{tab:matched}, the line ratings of each variant are multiplied by a common factor so that its worst-case threshold equals that of the reference model.
This lowers the ratings of the strongly damped variant by at most 1.1\% and raises those of the weakly damped one by at most 0.6\%, which works slightly against the effect we report.
Measured against the remaining margin of the most heavily loaded line, the changes are at most 5\%.
Without the rescaling, with the original ratings and each variant's kick energy measured in units of its own worst-case energy, halving instead of doubling the damping raises the risk of frequency kicks by similar factors, 1.4 to 1.8, where 10 to 50\% of kicks fail.

\paragraph*{Kicks at a single generator.}
For these kicks the worst case is an angle displacement that overloads a line at once, so it is identical for all variants without any rescaling.
The linear prediction remains close to the nonlinear simulations, though less close than for kicks spread over all generators: the largest differences are 0.07 for the IEEE grid and 0.045 for RTS-GMLC, because a single generator absorbs the whole kick and swings further.
At kick energies where 10 to 50\% of disturbances cause failure, halving instead of doubling the damping raises the risk by a factor of 1.2 to 3.2, and the mean reach orders the four variants correctly at all eight kick energies we compared on the two grids.

\paragraph*{Numerics.}
The response vectors of Eq.~\eqref{eq:r} are computed from the linearized dynamics on a time grid of 0.01~s over 10~s.
The nonlinear failure threshold of each direction is located by 30 bisection steps, each a fourth-order Runge--Kutta simulation with step 0.01~s that stops as soon as a line exceeds its rating.
Each curve uses the same 4096 random directions.
Halving the time step to 0.005 and then 0.0025~s changes the linear curves by less than 0.001 and the nonlinear thresholds by less than 0.2\%.
The bisection assumes that a direction that fails at some kick size fails at all larger ones.
We checked this for 512 directions at 80 kick sizes, in both grids and for both kick types, and found no exception.
At the kick sizes where 10, 30, and 50\% of kicks fail, the nonlinear and linear curves differ by between $-0.007$ and $+0.010$, with 95\% confidence half-widths below 0.007 from the paired directions.

\paragraph*{First-swing and later-swing failures.}
The end of the first swing, $t_1$, is the first minimum of $|\bm r_b(t)|$ after its first maximum, for the line end $b$ that holds the worst case in the reference model with frequency kicks.
The same $t_1$ is used for all variants of a grid and for both kick types.
The linear failure time of direction $u$ at kick size $B$ is the first time at which $\max_b \bm r_b(t)\cdot u\ge 1/B$.
The nonlinear failure time is the first time at which a line exceeds its rating in a single simulation at kick size $B$, with the same 4096 directions as above.
We compute it at seven kick energies, those at which 2, 5, 10, 20, 30, 40, and 50\% of kicks fail in the linear prediction for the reference model.
The ratio $\rho$ is the largest $|\bm r_b(t)|/G$ over all line ends and all $t>t_1$.
For frequency kicks, the first-swing failures of the three damping values agree to within 0.011, and moving $t_1$ by 20\% in either direction keeps them within 0.035.

\paragraph*{Overloads that must last.}
For the criterion of Sec.~\ref{sec:duration}, the linear reach is $g_\tau(u)=\max_{b,t}\,\min_{t\le s\le t+\tau}\bm r_b(s)\cdot u$, and a kick fails if $B\,g_\tau(u)\ge1$.
In the nonlinear simulations a kick fails if some line end exceeds its rating at every time step of an interval of length $\tau$; the failure time is the start of that interval.

\section{Geometry of the linear prediction}
\label{app:geometry}

Here we collect the formulas behind the three rules of Sec.~\ref{sec:hull} and the damping and inertia laws of Sec.~\ref{sec:grids}.
Directions $u$ are uniformly distributed on the unit sphere in $n$ dimensions.

\paragraph*{Single cap.}
For a single response vector of length $G$, the directions with $G\,u\cdot\hat{\bm r}>1/B$ form a cap whose area fraction is
\begin{equation}
  C_n(s) = \tfrac12\, I_{1-s^2}\!\left(\tfrac{n-1}{2},\tfrac12\right),\qquad s=\frac{1}{BG},
  \label{eq:cap}
\end{equation}
where $I$ is the regularized incomplete beta function.
Near onset, $s\to1$, the cap shrinks as $(1-s)^{(n-1)/2}$, which is why a worst-case cap is tiny in high dimensions.

\paragraph*{Turning.}
Let a response peak at time $t_*$ with height $G$, with $|\bm r(t)|\approx G[1-\kappa(t-t_*)^2/2]$ and with its direction turning at angular speed $\omega$.
Expanding $\bm r(t)\cdot u$ to second order around the peak and maximizing over $t$ shows that the dangerous directions near onset fill an ellipse instead of a circle, stretched along the direction of turning by $\sqrt{1+\Lambda^2}$ with $\Lambda=\omega/\sqrt{\kappa}$.
Summing over separate peaks $k$ gives
\begin{equation}
  \Plin(B)\approx\sum_k\sqrt{1+\Lambda_k^2}\;C_n\!\left(\frac{1}{BG_k}\right).
  \label{eq:onset}
\end{equation}
This is the Laplace approximation familiar from second-order reliability methods and from the extremes of Gaussian processes~\cite{Breitung1984,Piterbarg1996}.
Away from onset, summing the expected number of times $\bm r(t)\cdot u$ crosses the level $1/B$ (Rice's formula~\cite{Rice1944}) extends Eq.~\eqref{eq:onset} until directions begin to cross more than once.
For a curve of response vectors, the dangerous set is a tube around the curve on the sphere, whose volume was first computed by Hotelling~\cite{Hotelling1939}.

\paragraph*{Mean reach and effective number of directions.}
Because $g(u)$ is the support function of the hull, its average over directions is half the mean width of the hull.
For Gaussian kicks $\xi$ with independent unit components, $g(\xi)=|\xi|\,g(\xi/|\xi|)$, so the mean is simply multiplied by $\langle|\xi|\rangle\approx\sqrt{n}$; the mean of $g(\xi)$ is known as the Gaussian width of the hull~\cite{Vershynin2018}.
For Gaussian kicks, $g(\xi)$ is the supremum of the Gaussian process $X_{b,t}=\bm r_b(t)\cdot\xi$, together with the constant $X_0=0$ that represents the origin.
Its increments satisfy $\langle(X_{b,t}-X_{b',t'})^2\rangle=|\bm r_b(t)-\bm r_{b'}(t')|^2$.
Suppose every such distance, including every length $|\bm r_b(t)|$, is at least as large for one set of response vectors as for another.
The Sudakov--Fernique inequality~\cite{Vershynin2018} then states that the first set has at least as large a Gaussian width, and hence at least as large a mean reach.
This is the precise sense in which a response that turns and spreads out raises the mean reach.
We define the effective number of independent dangerous directions $N_{\rm eff}$ by requiring that $N_{\rm eff}$ mutually perpendicular response vectors of length $G$ have the same Gaussian width, that is, $\langle\max(0,Z_1,\dots,Z_{N_{\rm eff}})\rangle=\langle g(\xi)\rangle/G$ for independent standard normal $Z_k$.

\paragraph*{Damping.}
Without governors, the linearized swing equations read $M\ddot\delta+\gamma M\dot\delta+K\delta=0$, with $M$ the diagonal matrix of $2H_i/\omega_s$, $K$ the synchronizing matrix, and the same damping rate $\gamma$ for every machine.
Substituting $\delta=e^{-\gamma t/2}y$ gives $M\ddot y+(K-\gamma^2M/4)\,y=0$, with $y(0)=\delta(0)$ and $\dot y(0)=\dot\delta(0)+\gamma\delta(0)/2$.
For frequency kicks, $\delta(0)=0$, so $y$ is the undamped response with $K$ shifted by $\gamma^2M/4$, a relative change of order $(\gamma/2\omega)^2$.
Line flows are linear in $\delta$, so $\bm r_b(t;\gamma)\approx e^{-\gamma t/2}\bm r_b(t;0)$.
Because $g(u)=\bm r_{b_*}(t_*)\cdot u$ at its maximizer, Danskin's theorem (the envelope theorem for a maximum) gives $\partial g/\partial\gamma=\partial_\gamma\bm r_{b_*}(t_*)\cdot u=-t_*\,g/2$, which is Eq.~\eqref{eq:tlaw}.
In the full models, the exact derivative $\partial\ln g/\partial\ln D$, plotted against $t_*$ across directions, has slope $-0.25$~s$^{-1}$ in the IEEE grid, as predicted by $-\bar\gamma/2$; in RTS-GMLC the slope is $-0.24$~s$^{-1}$ against $-0.29$~s$^{-1}$ predicted.

For Eqs.~\eqref{eq:qexact} and \eqref{eq:qlaw}, let $a_p$ be the reach exceeded by a fraction $p$ of directions, ${\rm Prob}[g(u)>a_p]=p$, so that $E_p\propto a_p^{-2}$.
If $g$ depends smoothly on a parameter $\theta$ and has a continuous density at $a_p$, differentiating at fixed $p$ gives $da_p/d\theta=\langle\partial g/\partial\theta\mid g=a_p\rangle$~\cite{Hong2009}: the threshold moves with the average change of the directions that lie on it.
With $\theta=\ln D$, and because $g=a_p$ on the threshold, $d\ln a_p/d\ln D=-\langle s\rangle_p$.
Since $E_p/E_{\rm on}=(G/a_p)^2$ and $d\ln G/d\ln D=-s_G$, Eq.~\eqref{eq:qexact} follows.
The line ratings, and hence the margins in Eq.~\eqref{eq:r}, are held fixed.
For a common damping rate $\gamma$, scaling $D$ scales $\gamma$, and Eq.~\eqref{eq:tlaw} gives $s(u)=\gamma\,t_*(u)/2$ and $s_G=\gamma\,t_G/2$, hence Eq.~\eqref{eq:qlaw}.
With unequal rates $\gamma_i$, scaling $D$ scales every $\gamma_i$ by the same factor, and replacing the rates by their arithmetic mean $\bar\gamma$ is an approximation.
In the full models, $\bar\gamma=0.50$~s$^{-1}$ in the IEEE grid, where the rates range from 0.40 to 0.58~s$^{-1}$, and 0.58~s$^{-1}$ in RTS-GMLC, where they range from 0.40 to 0.71~s$^{-1}$.
Numerically, $s(u)$ is a central difference with $D$ scaled by $1\pm0.1$, and $\langle s\rangle_p$ and $\langle t_*\rangle_p$ are means over the directions whose reach lies within $\pm0.01$ of the level $p$ in probability, among $4\times10^4$ directions.

\paragraph*{Inertia.}
For power-impulse kicks, $\delta(0)=0$ and $M\dot\delta(0)=J$.
Multiply every inertia by $\lambda$, so that $M\to\lambda M$, and every damping rate by $\lambda^{-1/2}$.
Substituting $\delta_\lambda(t)=\lambda^{-1/2}\,\delta(t/\sqrt\lambda)$ into $\lambda M\ddot\delta_\lambda+\lambda^{-1/2}\gamma\,\lambda M\dot\delta_\lambda+K\delta_\lambda=0$ recovers the original equation, and $\lambda M\dot\delta_\lambda(0)=M\dot\delta(0)=J$, so the same impulse produces this response.
The argument holds equally for unequal damping rates, each scaled by $\lambda^{-1/2}$.
The operating point, and hence the margins, do not depend on the inertia, so $\bm r_b(t;\lambda)=\lambda^{-1/2}\,\bm r_b(t/\sqrt\lambda;1)$: the hull over a window of length $T$ is the old hull over the window $T/\sqrt\lambda$, multiplied by $\lambda^{-1/2}$.
With all inertias halved and no governors, the reach of every direction grows by a factor between 1.410 and 1.417 over the fixed window, and by exactly $\sqrt2$ when the window is rescaled as well.
If the damping rate, rather than the per-unit damping, is kept fixed when the inertia is halved, the later swings are damped less than the scaling law requires, and the mean reach grows more than the reach (1.50 against 1.43 in the IEEE grid).
Rescaling by the ratio of the worst cases then reproduces the low-inertia curve only to within 0.072 (IEEE) and 0.062 (RTS-GMLC).

\paragraph*{Spread.}
Because the hull lies inside a ball of radius $G$, the function $g(u)$ changes by at most $G$ times the distance between two directions.
The concentration of measure on the sphere then gives~\cite{Ledoux2001}
\begin{equation}
  {\rm Prob}\big[\,|g(u)-\langle g\rangle|\ge\epsilon\,\big] \le 2\exp\!\left[-\frac{(n-1)\,\epsilon^2}{2G^2}\right],
\end{equation}
which bounds the width of the rise of $\Plin(B)$.
The bound holds in all our examples but is about 3 to 6 times wider than the actual rise.
An exact expression for Gaussian kicks~\cite{Chatterjee2014} explains why:
\begin{equation}
  {\rm Var}\,g(\xi)=\int_0^1\big\langle \bm x_*(\xi)\cdot\bm x_*(\xi_\rho)\big\rangle\,d\rho,
\end{equation}
where $\xi_\rho=\rho\,\xi+\sqrt{1-\rho^2}\,\xi'$ is a kick correlated with $\xi$, and $\bm x_*(\xi)$ is the response vector that attains the maximum in Eq.~\eqref{eq:g}, that is, the failure route of $\xi$.
The variance is the average overlap of the failure routes of correlated kicks.
It is large when correlated kicks fail through the same route and small when they fail through different routes.
It never reaches the upper bound $G^2$ set by the Gaussian Poincar\'e inequality: even for a hull spanned by a single vector of length $G$ and the origin, $g=G\max(0,Z)$ with $Z$ standard normal, and the variance is $(1/2-1/2\pi)\,G^2\approx0.34\,G^2$.
We verified the identity numerically in all our examples, to within 2.5\%.

\paragraph*{Risk ordering.}
If $\mathcal H_1\subseteq\mathcal H_2$, then $g_1(u)\le g_2(u)$ for every $u$, so every kick that fails in system 1 also fails in system 2, whatever the distribution of kick directions.
If $\bm r_b(t;\gamma)=e^{-\gamma t/2}\,\bm r_b(t;0)$, a larger $\gamma$ multiplies every response vector by a factor between 0 and 1; since the origin belongs to the hull, the new hull lies inside the old one.
The mean reach does not order risk in this way.
In two dimensions, let $\mathcal H_A$ be the segment from $-\bm e_1$ to $\bm e_1$, and $\mathcal H_B$ the convex hull of $\bm e_1$ and the disk of radius 0.7 around the origin.
Both have reach 1.
Their reaches in the direction at angle $\theta$ are $g_A=|\cos\theta|$ and $g_B=\max(0.7,\cos\theta)$, with mean reaches $2/\pi\approx0.64$ and 0.75.
Yet at $1/B=0.9$ a fraction 0.29 of directions fails in A and 0.14 in B, while at $1/B=0.5$ the fractions are 0.67 and 1.
The curves cross, and the hull with the larger mean reach is the safer one near onset.

\section{The 12-dimensional networks}
\label{app:networks}

\paragraph*{Networks.}
All networks obey $\dot x=Ax+\beta\,x\circ x$ with 12 variables, where $(x\circ x)_i=x_i^2$.
Every variable decays at unit rate, so all eigenvalues of $A$ equal $-1$, and the output variable $x_1$ is driven by feed-forward chains.
In network A a single chain of three variables feeds $x_1$, each link with coupling $c_A=3.17$.
In network B three identical chains of three variables each feed $x_1$, with coupling $c_B=2.62$.
The couplings are chosen so that the longest response vector has the same length, $G=7.5$, in both networks.
Networks C and D have chains of lengths (2,4,5) and (3,8), respectively.
The coupling of each chain is chosen so that every chain alone produces the same peak response, and all couplings are then scaled together to give $G=7.5$.
Because their chains peak at different times, their response vectors turn and their hulls are wider than A's.
Their linear failure probabilities exceed A's by 0.12 and 0.18 on average over kick sizes $0.2\le B\le5$.

\paragraph*{Simulations.}
A kick sets $x(0)=Bu$ with $u$ uniform on the unit sphere, and the network fails if $x_1\ge1$ within $T=12$.
The failure threshold of each of 4096 directions is found by 40 bisection steps on $0\le B\le5$, using fourth-order Runge--Kutta integration with step 0.01.

\paragraph*{Rotated quadratic couplings.}
For an orthogonal matrix $Q$ that leaves the output coordinate unchanged, the network $\dot y=QAQ^{\rm T}y+\beta\,y\circ y$ has the same linear dynamics as the original, written in rotated coordinates.
Its response vectors are the original ones rotated by $Q$, so its response hull and its linear prediction are the same.
It is not, however, the original network in new coordinates, which would read $\dot y=QAQ^{\rm T}y+\beta\,Q\big[(Q^{\rm T}y)\circ(Q^{\rm T}y)\big]$.
Each $Q$ defines a different nonlinear system with the same linearization: only the quadratic coupling differs.
In the frame in which B's three chains are merged, $Q$ combines corresponding variables of the three chains into their sum and two differences.
The other frames use random $Q$.
We repeated all comparisons for $\beta=0.05$, 0.1, and 0.2 with three random frames each.
At $\beta=0.05$ the difference between B and A shrinks to what the linear prediction gives, and networks C and D, whose hulls are wider than A's, are riskier than A in every frame we tried, as their wider hulls predict.

\paragraph*{Second-order response.}
Writing $x=B\,x^{(1)}+B^2x^{(2)}+\dots$ gives $\dot x^{(1)}=Ax^{(1)}$ with $x^{(1)}(0)=u$, and $\dot x^{(2)}=Ax^{(2)}+\beta\,x^{(1)}\circ x^{(1)}$ with $x^{(2)}(0)=0$.
The second-order threshold of direction $u$ is the smallest $B$ for which $B\,x^{(1)}_1(t)+B^2x^{(2)}_1(t)\ge1$ at some time $t\le T$.
Both equations are linear in their unknowns and are integrated once per direction, which costs far less than locating the nonlinear threshold by repeated simulation.

\bibliography{references}

\end{document}